\documentclass[aps,preprintnumbers,superscriptaddress,showpacs,twocolumn]{revtex4}
\usepackage{epsfig}
\usepackage{psfrag}
\usepackage{amsfonts}
\usepackage{graphicx}
\usepackage{dcolumn}
\usepackage{bm}
\def\ie{ \emph{i.e.,} }

\begin{document}

\title{Electromagnetic responses of  a nonextensive quark-gluon plasma}

\author{Bing-feng Jiang}
\email{jiangbf@mails.ccnu.edu.cn} \affiliation{ College of Intelligent Systems Science and Engineering, Hubei Minzu University, Enshi 445000,
People's Republic of China }

\author{Jun Chen}
\affiliation{ College of Intelligent Systems Science and Engineering, Hubei Minzu University, Enshi 445000,
People's Republic of China }

\author{De-fu Hou}
\email{houdf@mail.ccnu.edu.cn} \affiliation{Key Laboratory of Quark and Lepton Physics
(MOE) and Institute
of Particle Physics, Central China Normal
University,Wuhan 430079,People's Republic of China}


\date{\today}
\begin{abstract}
Based on  the nonextensive statistical mechanics and the gluon polarization tensor obtained from kinetic theory,
we derive the longitudinal and transverse gluon self-energies  for the quark-gluon plasma(QGP).
The electric permittivity $\varepsilon$ and  the magnetic permeability $\mu_M$ are evaluated  from the gluon self-energies through which the real part of the square of the refraction index ${\rm Re}\, n^2$ and the Depine-Lakhtakia index $n_{DL}$ are investigated. The real part of $\varepsilon$ displays a frequency pole  $\omega_d=p$, which is just the position of the frequency inflexion of the imaginary part of $\varepsilon$.
The nonextensive parameter $q$ significantly affects the real and imaginary parts of $\varepsilon$ in the spacelike region $\omega<p$, while the frequency pole $\omega_d=p$ remains unchanged as $q$ increases. The magnetic permeability, ${\rm Re}\, n^2$ and the Depine-Lakhtakia index $n_{DL}$ diverge at frequency $\omega_m$. As $q$ increases, the pole frequency $\omega_m$ shifts to large frequency region. The Depine-Lakhtakia index $n_{DL}$ becomes negative in a quite large frequency region $\omega\in[\omega_c, \omega_m]$. When $q$ increases, the frequency range for $n_{DL}<0$ becomes wider.
Nevertheless, there are no propagating modes for the negative refraction.
In addition, as momentum $p$ increases, the electric permittivity, the magnetic permeability, ${\rm Re}\, n^2$ and $n_{DL}$ are sensitive to the change of $p$, which indicates the importance of the spatial dispersion in the electromagnetic responses of the QGP.
\end{abstract}

\maketitle
\section{Introduction}

Researchers in heavy ion community have expected to seek a special state of matter---quark-gluon plasma(QGP) predicted in quantum chromodynamics(QCD) by
colliding two heavy nuclei in terrestrial laboratories.
Some groups have applied viscous hydrodynamics to simulate the evolution of the hot dense nuclear matter produced in relativistic heavy ion collisions. The hydrodynamic simulations successfully reproduce the observables at RHIC and LHC, such as the elliptic flow, the particle
spectra\cite{heinz10,teaney,romatschke10,gale13,heller18}, etc.
Later, viscous effects on different aspects of QGP, such as dilepton production\cite{dulsing08},
jet quenching\cite{majumder07,dusling10,jiang15,sarkar18} and the related wakes\cite{bouras09,neufeld08,jiang12,jiang11},
heavy quarkonia dissociation\cite{noronha09}, heavy quark transport coefficients\cite{shaikh},
chiral magnetic effect\cite{jiang18,hidaka18,huang18},  quark polarization\cite{huang11} and the related anomalous transport phenomena\cite{fu21,shuai21,becattini21}, phase transition dynamics\cite{feng18} and dielectric properties\cite{jiang13,jiang10},
have been extensively addressed in recent years.

However, the hadron spectra  in heavy ion collisions are well described by
hydrodynamic model at low transverse momentum, especially in peripheral collisions\cite{strickland22epjc}. At high transverse momentum, hydrodynamic simulations predict an exponential behavior, while the experimental data  show  a power-law tail. It  is argued that the difference in hadron spectra between the hydrodynamic simulations and the experimental data at high transverse momentum is attributed to the thermal distributions, asymptotically of exponential in nature, adopted at freeze-out in hydrodynamic simulations\cite{strickland22epjc}.

There has been a great achievement in describing the observables for all considered momentum regions in relativistic heavy ion collisions by applying nonextensive statistics in recent years.  The nonextensive statistics has reproduced the hadron spectra the exponential form at low transverse momentum and the power-law tail at high transverse momentum perfectly. It is impressive that the nonextensive statistics can describe the experimental data of spectra covering 14 or 15 orders of magnitude from the lowest to the considered highest transverse momentum by using only three input parameters\cite{wong,zheng0,kapusta21}.

The Boltzmann-Gibbs statistics, which has long been used as a basic and standard framework for thermal physical systems, is inapplicable to systems with long range interactions and correlations, long-time memories, and fractal space-time structure\cite{strongint4,pre}. The nonextensive statistics has been proposed to apply to such systems\cite{tsallis,tsallis09}. In nonextensive statistics, entropy is not an additive quantity. And $q$ is the nonextensive parameter, the deviation of $q$ from unit characters the extent of the nonextensivity of a system. Systems subject to long range interactions and correlations will
demonstrate  nonextensive particle distributions which are away from the equilibrated thermal distribution\cite{pre}.
It is reported that the distribution of particles in space plasmas and laboratory plasmas in some cases are indeed not exactly Maxwellian distribution\cite{nonthermal1,nonthermal2,nonthermal3}.

In the context of relativistic heavy ion collisions,
the  temperature at RHIC and LHC reaches several times of the critical temperature $T_c$ of transition from the QGP phase to the hadronic phase and vice versa. At such  temperature, quarks  will be liberated from the hadron bags, but residual color-Coulomb interaction between constituents leads to a strongly interacting QGP\cite{strongint1,strongint2,strongint4,peshier}, where memory effects may not be negligible\cite{strongint4}. In addition, color long-range interaction will cause non-Markovian processes\cite{strongint4}. It is argued in Ref.\cite{mitra} that there are two-particle long range correlations in heavy ion collisions which result in ridge structure in particle multiplicity distributions.
In view of the facts mentioned above, it is a natural conjecture that the nonextensive statistics becomes involved in describing the QGP properties.

Though the study on implication of the nonextensive statistics in relativistic heavy ion collisions has become one of  hot topics in recent years, the aims  have mainly been  focused on the nonextensive effects on transverse momentum spectra, radial flow, the EOS of hadronic matter and the QGP, the transition from hadron phase to quark matter phase etc\cite{wong,zheng0,biro2009,plb2024,wong13,tang,cleymans13,liu14,zheng1,zheng2,cleymans15,cleymans16,
azmi,wong2,azmi20,rath20j,biro20,zhen21,zhen22,cleymansprd,kyan,zhao2020,zheng24,cleymans12epja,
cleymans12jpg,biro05prl,cleymans18epja,shen19epja,osada,biroepja2012,biro12prc,tsallis2022,prd94094026,epja53102}.
It is of great significance to apply nonextensive distribution functions in the dynamic models to study the properties of the QGP. There were a number of works that have been carried out along that line.
Some groups have derived  hydrodynamic equations  according to the nonextensive distributions and the kinetic theory\cite{osada,biroepja2012,biro12prc}. It is found that there is a correspondence between the ideal nonextensive  hydrodynamics and the usual dissipative hydrodynamics\cite{osada}. Then, some researchers have investigated the shear viscosity, bulk viscosity and other transport coefficients according to the nonextensive hydrodynamics and studied the nonextensive effects on them\cite{osada,biroepja2012,biro12prc}.
Alqahtani et al. have developed an anisotropic nonextensive hydrodynamics and studied the time evolution of fluid equations. They found that the nonextensive effects significantly affected the bulk pressure evolution\cite{strickland22epjc}.
In addition, the propagation of the nonlinear wave induced by  perturbations in QGP has been investigated  with the relativistic second order ideal and dissipative
hydrodynamics with a nonextensive equation of state\cite{rath20e,sarwar}. Within the framework of kinetic theory associated with the nonextensive distribution functions, Rath et al.  have investigated the nonextensive effects on the transport coefficients and the viscous properties of a magnetized hot and dense QCD matter\cite{rath23e,rath24}.
Meanwhile, some authors have studied the energy loss, jet quenching parameter $\hat{q}$ and the nuclear modification factor $R_{AA}$ in the QGP by using the nonextensive statistics\cite{cleymans16a,physa2023,deppman24}.
In spite of the above mentioned achievements, the study of the QGP properties in terms of the dynamical model combined with the nonextensive statistics should be investigated further.

In the present paper, we investigate the nonextensive effects on the electric permittivity $\varepsilon$, the magnetic permeability $\mu_M$ and the refractive properties of the QGP.
By applying the nonextensive distribution functions to the polarization tensor formula derived from the QGP kinetic theory\cite{thoma2000}, we can obtain the longitudinal and transverse gluon self-energies, through which the electric permittivity, the magnetic permeability, the square of the refractive index $n^2$ and the Depine-Lakhtakia index
$n_{DL}$  can be derived. Through the sequent derivation, the nonextensive parameter $q$ is embedded in the expressions of these electromagnetic quantities of the QGP. Therefore, one can study the effects of nonextensivity on them.

The paper is organized as follows. In Sec. \ref{em}, we  briefly review the formulism for the electromagnetic responses in a plasma.
In Sec. \ref{neem}, in terms of the nonextensive distribution functions and the polarization tensor obtained from the QGP kinetic theory, we  derive the electric permittivity and the magnetic permeability,  through which the square of the refractive index $n^2$ and the Depine-Lakhtakia index $n_{DL}$ can be determined. The nonextensive effects on the electromagnetic properties of the QGP are shown in Sec. \ref{numerical}. Section\ref{summary} is the summary and remarks.

The natural units $k_B=\hbar=c=1$, the metric $g_{\mu\nu}=(+,-,-,-)$ and  the following notations $P=(\omega,\textbf{p})$ are used in the paper.

\section{The electromagnetic responses in a plasma}\label{em}

In order to  covariantly investigate the electric and magnetic properties in a relativistic plasma, one should introduce a pair of four-vectors $\widetilde{E}^\mu$, $\widetilde{B}^\mu$ in terms of the fluid four-velocity $u^\nu$
\begin{equation}\label{f1}
\widetilde{E}^\mu=u_\nu F^{\nu\mu}, \ \ \ \ \ \widetilde{B}^\mu=\frac{1}{2}\epsilon^{\nu\lambda\rho\mu}F_{\nu\lambda}u_\rho
\end{equation}
and
\begin{equation}\label{f2}
F^{\mu\nu}=u^\mu\widetilde{E}^\nu-\widetilde{E}^\mu u^\nu-\epsilon^{\mu\nu\lambda\rho}\widetilde{B}_\lambda u_\rho,
\end{equation}
where the Greek index $\mu$ is not confused with the magnetic permeability $\mu_M$.
According to Eqs.(\ref{f1}) and (\ref{f2}),
one can obtain the Fourier-transformed free action
\begin{equation}\label{s0}
S_0=-\frac{1}{2} \int \frac{d^4P}{(2\pi)^4} \{\widetilde{E}^\mu(P)\widetilde{E}_\mu(-P)-\widetilde{B}^\mu(P)\widetilde{B}_\mu(-P)\}.
\end{equation}

Considering the finite temperature, quantum correction, the corresponding action correction is
\begin{equation}\label{sint}
S_{int}=-\frac{1}{2} \int \frac{d^4P}{(2\pi)^4} A^\mu(-P)\Pi_{\mu\nu}(P)A^\nu(P),
\end{equation}
where $A^\mu(P)$ is vector boson field in momentum space, and $\Pi_{\mu\nu}(P)$ is polarization tensor which embodies the medium effects of plasma. In homogeneous and isotropic medium,the polarization tensor can be divided into longitudinal and transverse parts $\Pi_{\mu\nu}(P)=\Pi_L(P) P^L_{\mu\nu}(P)+\Pi_T(P) P^T_{\mu\nu}(P)$ with projectors  defined  as $P^T_{00}=P^T_{0i}=P^T_{i0}=0$, $P^T_{ij}=\delta^{ij}-\frac{p^ip^j}{p^2}$, $P^L_{\mu\nu}=\frac{p^\mu p^\nu}{P^2}-g^{\mu\nu}-P^T_{\mu\nu}$\cite{kapusta,bellac}. Thus, the effective action including medium effects is
\begin{equation}\label{seff}
S_{eff}=S_0+S_{int},
\end{equation}
which  also can be described as
\begin{eqnarray}\label{seff1}
S_{eff}&=&-\frac{1}{2} \int \frac{d^4P}{(2\pi)^4} [\varepsilon\widetilde{E}^\mu(P)\widetilde{E}_\mu(-P)\nonumber\\&-&
\frac{1}{\mu_M}\widetilde{B}^\mu(P)\widetilde{B}_\mu(-P)].
\end{eqnarray}
In Eq.(\ref{seff1}), $\varepsilon$ and $\mu_M$ represent the electric permittivity and the magnetic permeability respectively  which can describe the difference of the electric and magnetic properties of the vector field in the medium and  those in the vacuum.
According to  Eqs.(\ref{s0})(\ref{sint}) and (\ref{seff1}), one can obtain the electric permittivity and the magnetic permeability in plasma as following:
\begin{equation}\label{die}
\varepsilon(\omega,p)=1-\frac{\Pi_L(\omega,p)}{P^2},
\end{equation}
\begin{equation}\label{mag}
\frac{1}{\mu_M(\omega,p)}=1+\frac{P^2\Pi_T(\omega,p)-\omega^2\Pi_L(\omega,p)}{p^2P^2}.
\end{equation}

We have briefly reviewed the electromagnetic properties in a homogeneous and isotropic plasma, for the  detailed derivation, please refer to Refs.\cite{weldon,meng,wang,jiang13,jiang16}.

The refraction index is a square definition of
the electric  permittivity and the magnetic permeability as
\begin{equation}
n^2=\varepsilon(\omega,p)\mu_M(\omega,p).\label{n2}
\end{equation}
The simultaneous  change of positive  $\varepsilon$ and $\mu_M$ to negative $-\varepsilon$ and $-\mu_M$  does not change $n^2$.
Veselago proposed that it corresponds to the transformation of the refraction index from one branch $n=\sqrt{\varepsilon(\omega,p)\mu_M(\omega,p)}$ to the other $n=-\sqrt{\varepsilon(\omega,p)\mu_M(\omega,p)}$, i.e., the turn from the general refraction index to the negative one\cite{veselago}. The physical nature of the negative refraction is that the electromagnetic phase velocity propagates opposite to the energy flow\cite{veselago,pendry,smith,depine,ramakrishna}.
The electric  permittivity and the magnetic permeability are generally complex-valued functions of $\omega$ and $k$, such as $\varepsilon(\omega,p)=\varepsilon_r(\omega,p)+i \varepsilon_i(\omega,p)$, $\mu_M(\omega,p)=\mu_r(\omega,p)+i \mu_i(\omega,p)$, so does the refraction index $n$.
According to the phase velocity propagating antiparallel  to the energy flow, some authors have derived
the criterion for the negative refraction \cite{depine}
\begin{equation}
n_{DL}=\varepsilon_r |\mu_M|+\mu_r|\varepsilon|<0,\label{neff}
\end{equation}
where $n_{DL}$ is the called Depine-Lakhtakia index.
$n_{DL}<0$ implies ${\rm Re}\, n <0$, otherwise we will have a normal refraction index.
In recent years, there are a number of works which have addressed the refractive index in the QGP\cite{wang,jiang13,jiang16,hattori1,hattori2,jamal} as well as in strongly coupled and correlation systems\cite{amariti11a,amariti11b,amariti13,ge,zhang,phukon1,phukon2,forcellajhep,mahapatra,forcellaprb,carvalho,jing1,jing2}.

\section {The electric permittivity and the magnetic permeability: the nonextensive approach}\label{neem}

\subsection{The nonextensive distribution functions}

In nonextensive statistics,  in the case of the vanish chemical potential, the single particle distribution functions for fermions (quarks and antiquarks) and bosons (gluons)  in the QGP can be expressed as \cite{rath20e,sarwar,rahamanIJMPA}
\begin{equation}\label{fermi}
n_{q,\bar{q}}=\frac{1}{\left[1+(q-1)\beta E_k\right]^{\frac{q}{q-1}}+1},
\end{equation}
\begin{equation}\label{bose}
n_g=\frac{1}{\left[1+(q-1)\beta E_k\right]^{\frac{q}{q-1}}-1}.
\end{equation}
$E_k=\sqrt{\mathbf{k}^2+m^2}$ is single particle energy with mass $m$ and $\beta$ is related to temperature $\beta=\frac{1}{T}$.
The nonextensive distribution is also considered as a nonequilibrium distribution and the nonextensive parameter $q$ measures the extent of deviation from equilibrated thermal distribution\cite{strickland22epjc,osada,rath20e,sarwar,rath23e,rath24}. If $q=1$, the nonextentive distribution functions turn to the usual Fermi-Dirac distribution and Bose-Einstein  distribution, and the entropy recovers the usual  additive one.

We assume that the QGP system is at high temperature, the  nonextensive Fermi-Dirac distribution and Bose-Einstein  distribution Eqs.(\ref{fermi})(\ref{bose}) can approximate as\cite{biro2009,zheng1,cleymans16,biro05prl,cleymans18epja,shen19epja}
\begin{equation}\label{fer}
n_{q,\bar{q},g}={\left[1+(q-1)\beta E_k\right]^{-\frac{q}{q-1}}}.
\end{equation}
It is argued that these distribution functions can satisfy the consistency of thermodynamics\cite{biro2009,zheng1,cleymans16,cleymans12epja,cleymans12jpg,cleymans18epja,shen19epja}.
In addition, some other nonextensive distribution functions  $\frac{1}{\left[1+(q-1)\beta E_k\right]^{\frac{1}{q-1}}\pm1}$ and its high temperature approximation ${\left[1+(q-1)\beta E_k\right]^{-\frac{1}{q-1}}}$ which are slightly different from Eqs.(\ref{fermi})(\ref{bose}) and (\ref{fer}) are widely employed to study heavy ion phenomenology\cite{strickland22epjc,mitra,strongint4,sarwar,rath23e,rath24,rath20e}.

In relativistic heavy ion collisions, low transverse momentum hadrons show exponential decay in energy which are produced with the normal kinetic and chemical equilibrium. While high transverse momentum hadrons which are secondary showering products in hard scatterings exhibit power-law tail because of the asymptotic freedom of QCD\cite{kapusta21,wong}.
Kapusta argued that though the nonextensive single particle distribution functions have been not derived from QCD,
they can interpolate between an exponential at low transverse momentum, reflecting the thermal equilibrium, to a power law at  high transverse momentum, reflecting the asymptotic freedom of QCD\cite{kapusta21}.

By expanding Eq.(\ref{fer}) around $q=1$, one can get\cite{cleymans16}
\begin{eqnarray}\label{appnondis}
n_{q,\bar{q},g}=\textbf{e}^{-\beta E_k}+\frac{1}{2}(q-1)(-2+\beta E_k)\beta E_k\textbf{e}^{-\beta E_k}+...
~.\end{eqnarray}
When $q=1$, Eq.(\ref{appnondis}) leaves behind the first term which is the Boltzmann distribution exactly.

\subsection{The gluon polarization tensor in kinetic theory}

In recent years, many researchers have applied the transport theory and the hydrodynamic approach to study the nonextensive properties of the QGP \cite{strickland22epjc,strongint4,pre,mitra,biro2009,biro05prl,zheng24,osada,biroepja2012,biro12prc,rath20e,sarwar,rath23e,rath24,cleymans16a,physa2023,deppman24}. In addition, there are some attempts which have focused on constructing thermal field theory based on the nonextensive statistics\cite{rahamanIJMPA,kohyamaPTP,carvalhoPLB}.  In this section, we will apply the QGP kinetic theory associated with nonextensive distribution functions to derive the hard loop gluon polarization tensor with linear response method. For the detailed derivation, one  can  refer to Refs.\cite{thoma2000,schenke,strickland2003,guoyun2023,jamal}.

In the kinetic theory, the hard partons are described by their phase space densities, \ie the gauge covariant Wigner functions $Q^m(k,X)$, where $m$ denotes the  species of plasma particles,  quark, antiquark and gluon, $m\in{q, \overline{q},g}$.
$Q^m(k,X)$ are  $N_c\times N_c$  (for quark and antiquark) and $(N_c^2-1)\times (N_c^2-1)$ matrices (for gluon) in color space(for a $SU(N_c)$ color group). Usually, one can expand $Q^m(k,X)$ in terms of  color neutral background parts $Q^m_0(\textbf{k})$
and  color fluctuating parts $\delta Q^m(k,X)$ as  $Q^m(k,X)=Q^m_0(\textbf{k})+\delta Q^m(k,X)$ to construct linear transport equations.
The color neutral background parts are defined as $Q^{q/\overline{q}}_0(\textbf{k})=n^{q/\overline{q}}(\textbf{k})I$ and $Q^{g}_0(\textbf{k})=n^{g}(\textbf{k})\mathcal{I}$. The color fluctuations are denoted as $\delta Q^{q/\overline{q}}(k,X)=\delta n^{q/\overline{q}}_a(k,X)t^a$ and $\delta Q^g(k,X)=\delta n^g_a(k,X)T^a$.
$I$, $\mathcal{I}$ and $t^a$, $T^a$
are the unit matrices and  the generators in the fundamental and adjoint representations respectively. The  scalar functions $n^{q/\overline{q}}(\textbf{k})$, $n^{g}(\textbf{k})$, $\delta n^{q/\overline{q}}_a(k,X)$ and $\delta n^g_a(k,X)$ can be defined according to the Wigner functions $Q^m(k,X)$\cite{schenke,guoyun2023}
\begin{eqnarray}
n^{q/\overline{q}}(\textbf{k})=\frac{1}{N_c}{\rm Tr}[Q^{q/\overline{q}}(k,X)],  \\ \nonumber
n^{g}(\textbf{k})=\frac{1}{N_c^2-1}{\rm Tr}[Q^g(k,X)],
\end{eqnarray}
and
\begin{eqnarray}
\delta n^{q/\overline{q}}_a(k,X)=2{\rm Tr}[t_aQ^{q/\overline{q}}(k,X)], \\ \nonumber
\delta n^g_a(k,X)=\frac{1}{N_c}{\rm Tr}[T_aQ^g(k,X)].
\end{eqnarray}

We assume that the QGP experiences small fluctuations which is slightly away from stationary state $\delta n^m_a(k,X)<<n^m(\mathbf{k})$.
The linearized collisionless transport equations for the QGP can be given as\cite{thoma2000,schenke,strickland2003,guoyun2023,jiang16,jamal}
\begin{equation}
V\cdot \partial_X \delta n^m_a(k,X) + g \theta_m
v_{\mu}F^{\mu\nu}_a(X)\partial_{\nu}^{(k)}n^m(\mathbf{k})=0\label{trans2}\,\text{,}
\end{equation}
where $g$ is the strong coupling constant. $\theta_{g}=\theta_{q}=1$, $\theta_{\bar{q}}=-1$ and $\partial_{\nu}^{(k)}$ denotes the four-momentum derivative. The linearized field strength tensor is
$F^{\mu\nu}=\partial^{\mu}A^{\nu}-\partial^{\nu}A^{\mu}$   with the soft field
$A^{\mu}=A^{\mu}_a T^a$ or $\mathcal{A}^{\mu}=\mathcal{A}^{\mu}_a t^a$.
At high temperature, the masses of the medium particles are much less than the temperature. Thus the four momentum of medium particles can be expressed as $K=(k,\textbf{k})$ with $k=|\textbf{k}|$ and $V=(1,\frac{\textbf{k}}{k})$.

The total induced color current due to the fluctuations of $\delta n_a^
m(k,X)$ can be expressed as\cite{thoma2000,schenke,strickland2003,guoyun2023,jiang16,jamal}
\begin{eqnarray}
J_{\text{ind}\,a}^\mu(X)&=&g \int \frac{d^3k}{(2\pi)^3} v^{\mu}[2N_c \delta n^{g}_a(k,X)\\ \nonumber&+&N_{f}[\delta n^{q}_a(k,X)
-\delta n^{\bar{q}}_a(k,X)]].
\end{eqnarray}

Solving the transport equation (\ref{trans2}) and Fourier-transforming,  one can get
\begin{equation}
	\delta n^m_a(k,P)=\frac{-ig\theta_m v_{\mu}F_a^{\mu\nu}(P)\partial_{\nu}^{(k)}n^m(\mathbf{k})}
{P\cdot V+i0^+}.
\end{equation}
Substituting the  obtained solutions $\delta n^m_a(k,P)$ into the Fourier transformation total induced color current, according to the the relation $\Pi^{\mu\nu}_{ab}=\frac{\delta J_{\text{ind}\,a}^\mu(P)}{\delta A^b_\nu(P)}$, we can obtain the gluon polarization tensor $\Pi^{\mu\nu}_{ab}(P)$ as\cite{thoma2000,schenke,strickland2003,guoyun2023,jiang16,jamal}
\begin{equation}\label{poltensor0}
\Pi^{\mu\nu}_{ab}(P)=g^2\delta_{ab}\int \frac{d^3k}{(2\pi)^3} v^\mu \frac{\partial f(\textbf{k})}{\partial k_l} [g^{l\nu}-\frac{v^\nu p^l}{P\cdot V+i0^+}],
\end{equation}
where  $f(\textbf{k})$ relates to the distribution functions of quark, antiquark and gluon $n_q(\textbf{k}), n_{\bar{q}}(\textbf{k}),n_g(\textbf{k})$ as
\begin{equation}\label{dis}
f(\textbf{k})=2N_c n_g(\textbf{k})+N_f(n_q(\textbf{k})+n_{\bar{q}}(\textbf{k})).
\end{equation}
The spatial parts of the gluon polarization tensor read
\begin{equation}\label{poltensor}
\Pi^{ij}(P)=-g^2\int \frac{d^3k}{(2\pi)^3} v^i \frac{\partial f(\textbf{k})}{\partial k^l} [\delta^{lj}+\frac{p^lv^j}{\omega-\textbf{p}\cdot\textbf{\^v}+i0^+}].
\end{equation}

It should be noted  that  the gluon polarization tensor Eq.(\ref{poltensor0})
is viable for  a relativistic plasma either in equilibrium but anisotropy caused by external fields or out of equilibrium\cite{thoma2000}.
Some other authors also argued that the derivation of Eq.(\ref{poltensor0}) in kinetic theory is applicative to arbitrary distribution functions $f(\mathbf{k})$\cite{strickland2003,guoyun2023} so long as the space-time is homogeneous\cite{guoyun2023}(please refer to Ch. 2.3).
The hard loop gluon self-energies calculated from Eq.(\ref{poltensor0}) for both equilibrium and nonequilibrium QGP are exactly consistent with those calculated based on the QCD diagrammatic approach respectively\cite{thoma2000}. These facts indicate the reliability of the kinetic theory method applicable to  the QGP either in equilibrium or out of equilibrium.
It should be stressed that a plasma system with nonextensive distributions (\ref{fermi})(\ref{bose}) signifies  a stationary, homogenous and colorless plasma state\cite{osada,biro12prc}. We assume that the QGP experiences small fluctuations which is slightly away from that plasma state. It makes sense to apply the kinetic theory to study the properties of nonextensive quark gluon systems.

Recently, there are a  number of works  which have focused on the  anisotropic properties of the QGP by combining Eq.(\ref{poltensor}) and Eq.(\ref{dis}) with anisotropic distribution functions(please refer to references in reviews\cite{aniso1,aniso2}).
In addition, some people have employed Eq.(\ref{poltensor}) and Eq.(\ref{dis}) with a distribution function modified by bulk viscosity to study the  properties of heavy quark\cite{bvis1,bvis2,bvis3}.
It is of great interest to apply Eq.(\ref{poltensor}) and Eq.(\ref{dis}) associated with nonextensive distribution functions (\ref{fermi})(\ref{bose})
to study the QGP properties.

\subsection{The gluon self-energies and the electromagnetic properties of the QGP }

We assume that the nonextensive parameter $q$ is independent on momentum, if the expanded nonextensive distribution functions  Eq.(\ref{appnondis}) are employed in Eq.(\ref{dis}),  one can prove
\begin{equation}\label{diffdis}
\frac{\partial f(\textbf{k})}{\partial k^l}=\frac{k^l}{k}\frac{\partial f(\textbf{k})}{\partial k}=v^l\frac{\partial f(\textbf{k})}{\partial k},
\end{equation}
for the detailed derivation please refer to  Appendix A.

According to Eq.(\ref{diffdis}), one can decouple  Eq.(\ref{poltensor}) into  the integral over $k$ and the integral over the solid angle $\Omega$
\begin{equation}\label{poltensordeco}
\Pi^{ij}(P)=m_{D}^2\int \frac{d\Omega}{4\pi}v^iv^l \cdot [\delta^{lj}+\frac{p^lv^j}{\omega-\textbf{p}\cdot\textbf{\^v}+i0^+}].
\end{equation}
$m_{D}^2$ is the square of Debye mass which is denoted as
\begin{equation}\label{debyemass}
m_{D}^2=-\frac{g^2}{2\pi^2}\int_0^\infty k^2 dk \frac{\partial f(\textbf{k})}{\partial k}.
\end{equation}

Substituting  Eq.(\ref{appnondis}) into  Eqs.(\ref{dis})(\ref{debyemass}), one can get an analytical result for the Debye mass in the nonextensive case
\begin{equation}\label{debyemasse}
m_{DE}^2=\frac{2(1+(q-1))(N_c+N_f)}{\pi^2}g^2T^2.
\end{equation}
For the detailed derivation please refer to  Appendix B.
By using nonextensive distribution functions  without approximation, some authors have derived an expression for the square of the Debye mass which is a complicated polynomial, please refer to Eq.(72) in Ref.\cite{mitra} .
When the usual Fermi-Dirac and Bose-Einstein distribution functions are used in  $f(\textbf{k})$ in Eq.(\ref{debyemass}), one can obtain the familiar result of the square of Debye mass
\begin{equation}\label{debyemassHTL}
m_{DH}^2=\frac{(2N_c+N_f)}{6}g^2T^2.
\end{equation}

The gluon longitudinal and transverse self-energies can be derived in terms of $\Pi^{ij}(P)$ and the projectors
\begin{equation}\label{logselfenergy1}
\Pi_L(P)=-\frac{P^2}{p^2}\Pi^{00}(P)=\frac{P^2}{\omega^2}\cdot \frac{p^ip^j}{p^2}\cdot\Pi^{ij}(P),
\end{equation}
\begin{equation}\label{transelfenergy1}
\Pi_T(P)=\frac{1}{2}(\delta^{ij}-\frac{p^ip^j}{p^2})\Pi^{ij}(P).
\end{equation}
After some algebra, we can obtain the gluon longitudinal and the transverse  self-energies
\begin{widetext}
\begin{eqnarray}\label{logselfenergy}
\Pi_L(P)
=\frac{2(1+(q-1))(N_c+N_f)}{\pi^2}g^2T^2\cdot \frac{P^2}{p^2}\cdot [-1+\frac{\omega}{2p}(\ln|\frac{\omega+p}{\omega-p}|-i\pi\delta(p^2-\omega^2))],
\end{eqnarray}
\begin{eqnarray}\label{transelfenergyfinal}
\Pi_T(P)
=\frac{(1+(q-1))(N_c+N_f)}{\pi^2}g^2T^2 \cdot[\frac{\omega^2}{p^2} +\frac{\omega(p^2-\omega^2)}{2p^3}\cdot(\ln|\frac{\omega+p}{\omega-p}|-i\pi\delta(p^2-\omega^2))].
\end{eqnarray}
For the detailed derivation, please refer to Appendix C. Substituting the longitudinal and transverse self-energies Eqs.(\ref{logselfenergy})(\ref{transelfenergyfinal}) into Eqs.(\ref{die})(\ref{mag}), we can obtain the electric permittivity and the magnetic permeability
\begin{eqnarray}\label{logdie}
\varepsilon(\omega,p)
=1+\frac{2(1+(q-1))(N_c+N_f)}{\pi^2}g^2T^2\cdot \frac{1}{p^2}[1-\frac{\omega}{2p}(\ln|\frac{\omega+p}{\omega-p}|-i\pi\delta(p^2-\omega^2))],
\end{eqnarray}
\begin{eqnarray}\label{magpeafinal}
\mu_M(\omega,p)=\frac{4p^4}{4p^4+6m_{DE}^2\omega^2+
m_{DE}^2(p^2-3\omega^2)\frac{\omega}{p}(\ln|\frac{\omega+p}{\omega-p}|-i\pi\delta(p^2-\omega^2))}.
\end{eqnarray}
\end{widetext}

\begin{figure}[t]
\begin{minipage}[h]{0.48\textwidth}
\centering{\includegraphics[width=7.5cm,height=4.5cm]{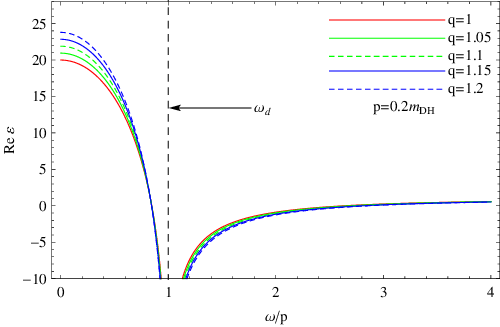}}
\end{minipage}
\begin{minipage}[h]{0.48\textwidth}
\centering{\includegraphics[width=7.5cm,height=4.5cm] {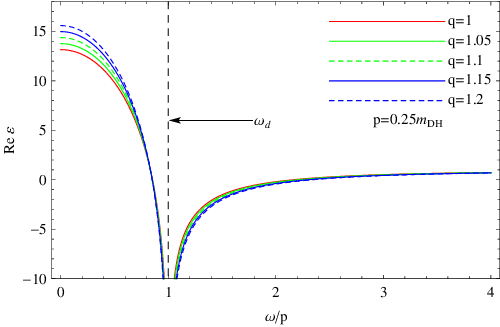}}
\end{minipage}
\begin{minipage}[h]{0.48\textwidth}
\centering{\includegraphics[width=7.5cm,height=4.5cm] {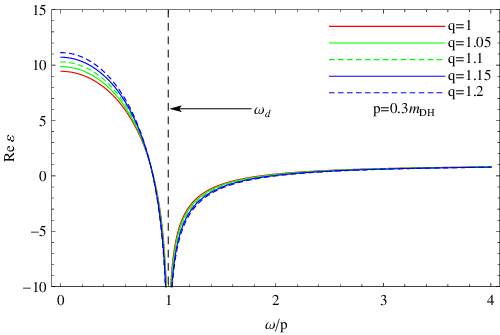}}
\end{minipage}
\caption{(color online) The real part of the electric permittivity of a nonextensive QGP. The red, green, dashed-green, blue and dashed-blue curves are for the cases of $q=1, 1.05, 1.1, 1.15, 1.2$ respectively. Top panel: $p=0.2m_{DH}$, Middle panel: $p=0.25m_{DH}$, Bottom panel: $p=0.3m_{DH}$. } \label{electric}
\end{figure}

\begin{figure}[t]
\begin{minipage}[h]{0.48\textwidth}
\centering{\includegraphics[width=7.5cm,height=4.635cm] {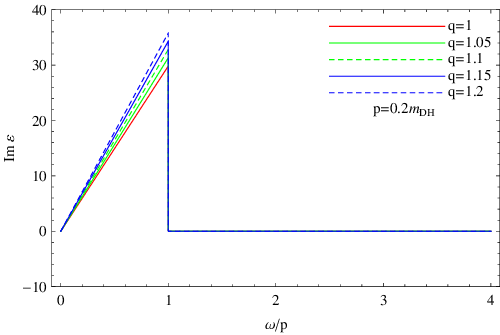}}
\end{minipage}
\begin{minipage}[h]{0.48\textwidth}
\centering{\includegraphics[width=7.5cm,height=4.5cm] {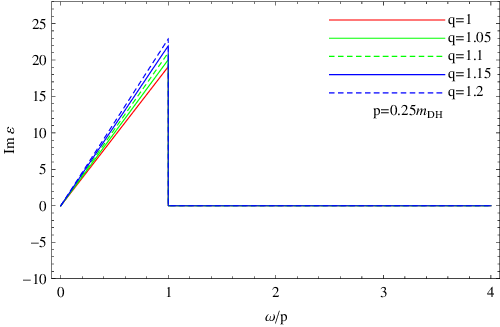}}
\end{minipage}
\begin{minipage}[h]{0.48\textwidth}
\centering{\includegraphics[width=7.5cm,height=4.5cm] {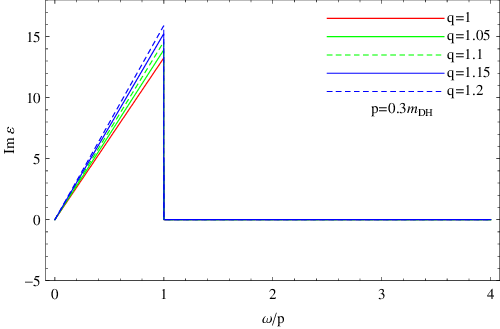}}
\end{minipage}
\caption{(color online) The imaginary part of the electric permittivity of a nonextensive QGP. The red, green, dashed-green, blue and dashed-blue curves are for the cases of $q=1, 1.05, 1.1, 1.15, 1.2$ respectively. Top panel: $p=0.2m_{DH}$, Middle panel: $p=0.25m_{DH}$, Bottom panel: $p=0.3m_{DH}$. } \label{magnetic}
\end{figure}

We have briefly reviewed the determination  of the electric permittivity and the magnetic permeability of the QGP
according to the polarization tensor derived from kinetic theory associated with the nonextensive distribution functions.  The nonextensive parameter $q$ is encoded in the gluon self-energies, the electric permittivity and the magnetic permeability. Combining with Eqs.~(\ref{logdie})(\ref{magpeafinal}) and (\ref{n2})(\ref{neff}), we can study the nonextensive effects on the  electromagnetic response properties of the QGP.

\begin{figure}[t]
\begin{minipage}[h]{0.48\textwidth}
\centering{\includegraphics[width=7.5cm,height=4.635cm] {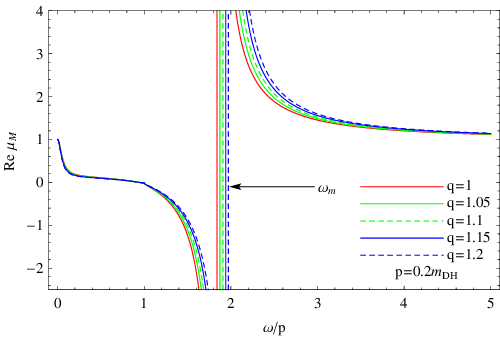}}
\end{minipage}
\begin{minipage}[h]{0.48\textwidth}
\centering{\includegraphics[width=7.5cm,height=4.5cm] {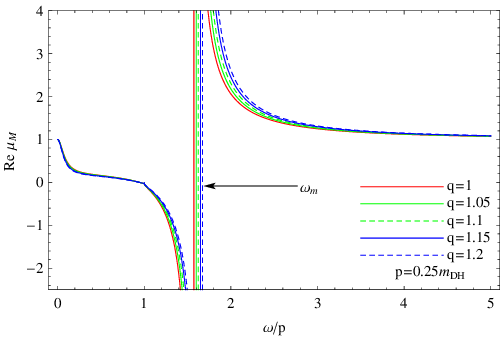}}
\end{minipage}
\begin{minipage}[h]{0.48\textwidth}
\centering{\includegraphics[width=7.5cm,height=4.5cm] {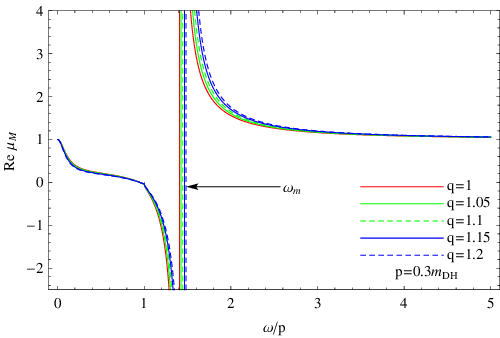}}
\end{minipage}
\caption{(color online) The real part of the magnetic permeability of a nonextensive QGP. The red, green, dashed-green, blue and dashed-blue curves are for the cases of $q=1, 1.05, 1.1, 1.15, 1.2$ respectively. Top panel: $p=0.2m_{DH}$, Middle panel: $p=0.25m_{DH}$, Bottom panel: $p=0.3m_{DH}$.} \label{magnetic}
\end{figure}

\begin{figure}[t]
\begin{minipage}[h]{0.48\textwidth}
\centering{\includegraphics[width=7.5cm,height=4.635cm] {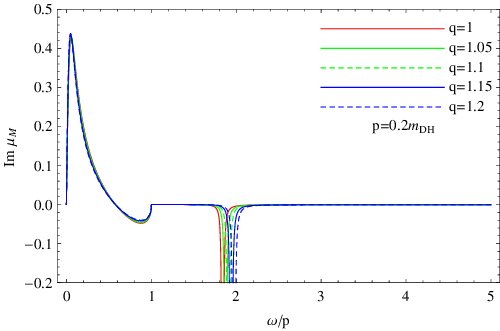}}
\end{minipage}
\begin{minipage}[h]{0.48\textwidth}
\centering{\includegraphics[width=7.5cm,height=4.5cm] {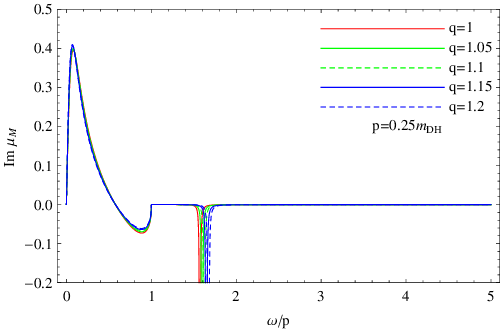}}
\end{minipage}
\begin{minipage}[h]{0.48\textwidth}
\centering{\includegraphics[width=7.5cm,height=4.5cm] {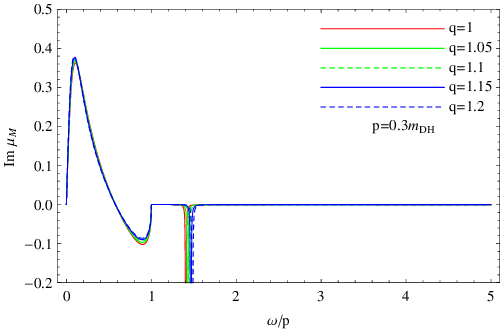}}
\end{minipage}
\caption{(color online) The imaginary part of the magnetic permeability of a nonextensive QGP. The red, green, dashed-green, blue and dashed-blue curves are for the cases of $q=1, 1.05, 1.1, 1.15, 1.2$ respectively. Top panel: $p=0.2m_{DH}$, Middle panel: $p=0.25m_{DH}$, Bottom panel: $p=0.3m_{DH}$.} \label{magnetic}
\end{figure}

\section{Numerical results}\label{numerical}

In the following, we treat the nonextensive parameter $q$ as an input parameter to study its effects on the electromagnetic properties of the QGP. Some investigations show that the consistency of the nonextensive
thermodynamics requires   $q>1$\cite{tsallis2022,osada,prd94094026} but $q<\frac{4}{3}$ \cite{prd94094026,epja53102}.
Researchers found to fit the RHIC and the LHC experimental data,  $q$ deviates from unit slightly. For example,
the parameter $q$ is in the range from $1.1$ to $1.2$ by fitting the transverse momentum distribution of the identical charged particles\cite{cleymans12jpg}. In Ref\cite{zheng0}, by fitting the spectra of particles with nonextensive distribution, the authors obtained the limits $q\in[1.063, 1.25]$   for mesons and  $q\in[1.05,1.25]$ for protons respectively. The most fitting values of $q$ from different groups lie in those ranges.
On the other hand, in view of the fact that the expanded nonextensive distribution functions Eq.(\ref{appnondis}) are adopted, the value of $q$ should not be far from $1$. Cleymans et al drew a conclusion that $q$ should deviate from unit by $20\%$ at most\cite{cleymans12jpg}.
Therefore, we will study the electromagnetic responses of the QGP with some explicit values of $q$ in range $q\in[1,1.2]$.
On the other hand, $N_c=3$ and $N_f=2$ are adopted to perform the numerical analysis.
In addition, such scaled momenta $p=0.2m_{DH},0.25m_{DH},0.3m_{DH}$  are used to study the $\omega$-dependent behavior of the electric permittivity, the magnetic permeability, the real part of the square of the refractive index ${\rm Re}\,  n^2$ and the Depine-Lakhtakia index $n_{DL}$.

We plot the real part of the electric permittivity  in Fig.1.  It is clear that there is a pole for ${\rm Re}\, \varepsilon$ at frequency  $\omega_{d}=p$.
In the spacelike region $\frac{\omega}{p}<1$, for a small $\omega$, the real part of the electric permittivity increases as the increase of the nonextensive parameter $q$. $q$ influences ${\rm Re}\, \varepsilon$  significantly.
As the increase of $\omega$, the difference between several nonextensive curves of ${\rm Re}\, \varepsilon$ becomes smaller and several  curves overlap each other for $\omega$ approaching $p$.
In the timelike region $\frac{\omega}{p}>1$, ${\rm Re}\, \varepsilon$ reduces as the increase of $q$.
$q$ has appreciated effects on ${\rm Re}\, \varepsilon$ for frequency around $\omega_{d}$.
As $\omega$ increases, all ${\rm Re}\, \varepsilon$ curves turn to unit asymptotically.
It should be noted that $q$ does not change the frequency position of the pole $\omega_{d}=p$. In addition, as shown from top to bottom panel, one can see ${\rm Re}\, \varepsilon$  significantly decreases as momentum increases from $0.2m_{DH}$  to $0.3m_{DH}$. At the frequency pole, the relation $\frac{\omega}{p}=1$ remains unchanged when $p$ varies.

\begin{figure}[t]
\begin{minipage}[h]{0.48\textwidth}
\centering{\includegraphics[width=7.5cm,height=4.635cm] {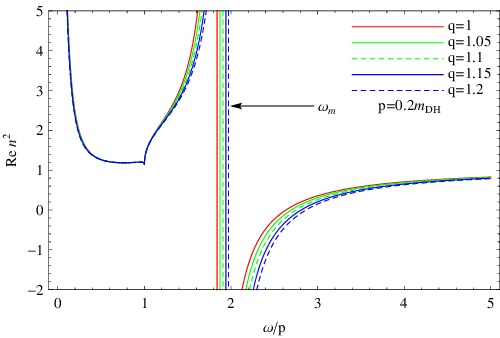}}
\end{minipage}
\begin{minipage}[h]{0.48\textwidth}
\centering{\includegraphics[width=7.5cm,height=4.5cm] {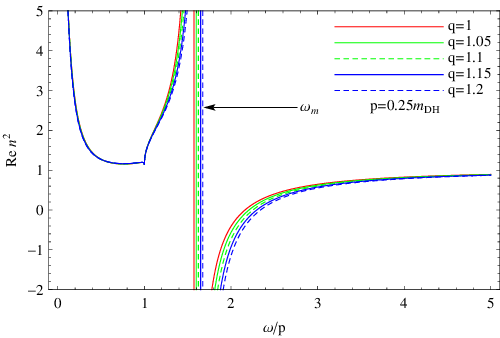}}
\end{minipage}
\begin{minipage}[h]{0.48\textwidth}
\centering{\includegraphics[width=7.5cm,height=4.5cm] {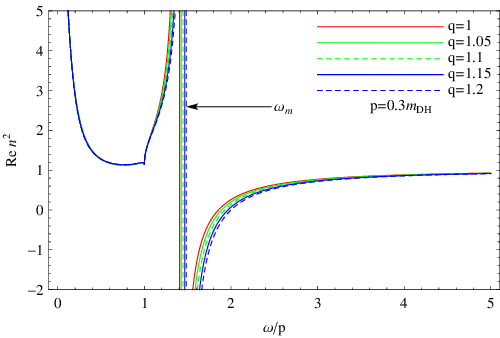}}
\end{minipage}
\caption{(color online) The real part of $n^2$ of a nonextensive QGP. The red, green, dashed-green, blue and dashed-blue curves are for the cases of $q=1, 1.05, 1.1, 1.15, 1.2$ respectively. Top panel: $p=0.2m_{DH}$, Middle panel: $p=0.25m_{DH}$, Bottom panel: $p=0.3m_{DH}$.} \label{magnetic}
\end{figure}

\begin{figure}
\begin{minipage}[h]{0.48\textwidth}
\centering{\includegraphics[width=7.5cm,height=4.635cm] {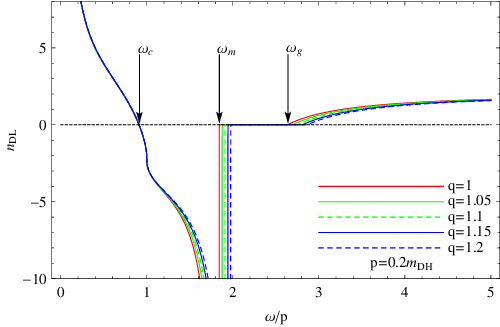}}
\end{minipage}
\begin{minipage}[h]{0.48\textwidth}
\centering{\includegraphics[width=7.5cm,height=4.5cm] {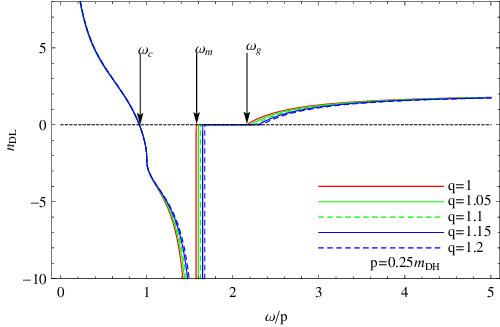}}
\end{minipage}
\begin{minipage}[h]{0.48\textwidth}
\centering{\includegraphics[width=7.5cm,height=4.5cm] {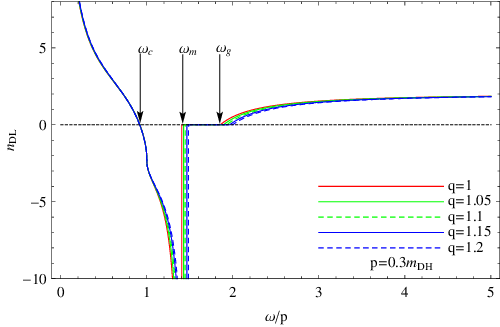}}
\end{minipage}
\caption{(color online) The Depine-Lakhtakia index $n_{DL}$ of a nonextensive QGP. The red, green, dashed-green, blue and dashed-blue curves are for the cases of $q=1, 1.05, 1.1, 1.15, 1.2$ respectively. Top panel: $p=0.2m_{DH}$, Middle panel: $p=0.25m_{DH}$, Bottom panel: $p=0.3m_{DH}$. } \label{magnetic}
\end{figure}
We present the  imaginary part of the electric permittivity in Fig.2. $\varepsilon$ has a nonzero imaginary part in the spacelike region $\frac{\omega}{p}<1$ which comes from Landau damping.
In the timelike region $\frac{\omega}{p}>1$,  ${\rm Im}\, \varepsilon=0$.
There is a frequency inflexion for ${\rm Im}\, \varepsilon$ located at $\omega_{d}=p$ which is just the pole frequency of ${\rm Re}\, \varepsilon$.
In the spacelike region $\frac{\omega}{p}<1$, the nonextensive parameter $q$ enhances the imaginary part of the electric permittivity. While in the timelike region $\frac{\omega}{p}>1$, ${\rm Im}\, \varepsilon$ remains zero  when $q$ changes.
In addition, $q$ does not change the inflexion frequency of ${\rm Im}\, \varepsilon$.
From top to  bottom panel in Fig.2,  one can see that as the real part,  ${\rm Im}\, \varepsilon$ significantly decreases  as momentum increases in the spacelike region.  But momentum $p$ does not change the relation $\frac{\omega}{p}=1$ at the frequency inflexion of ${\rm Im}\, \varepsilon$.

Fig.3 and Fig.4 display the real and imaginary parts of the magnetic permeability for different values of the nonextensive parameter. Both  the real and imaginary parts of the magnetic
permeability diverge at  frequency $\omega_m$.
It is easy to see that  $\omega_m$ shifts to large frequency region with the increase of $q$.
$\mu_M$ has a nonzero imaginary part in the frequency region $\frac{\omega}{p}<1$, but $q$ has a trivial impact on it.
In the same frequency region $\frac{\omega}{p}<1$, several q-curves of ${\rm Re}\, \mu_M$ coincide with each other, which indicates that $q$ has hardly any effect on ${\rm Re}\, \mu_M$ in that frequency reguion.
While in the frequency region $\frac{\omega}{p}>1$, $q$ has an important impact on ${\rm Re}\, \mu_M$ in frequency region $p<\omega<3p$.
When momentum $p$ increases from $0.2m_{DH}$  to $0.3m_{DH}$, pole frequency $\omega_m$  turns to small frequency values.

\begin{figure*}
\centerline{
\begin{minipage}[h]{0.48\textwidth}
\centering{\includegraphics[width=7.5cm,height=5cm] {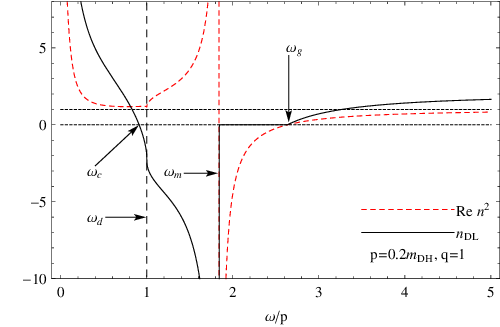}}
\end{minipage}
\begin{minipage}[h]{0.48\textwidth}
\centering{\includegraphics[width=7.5cm,height=5cm] {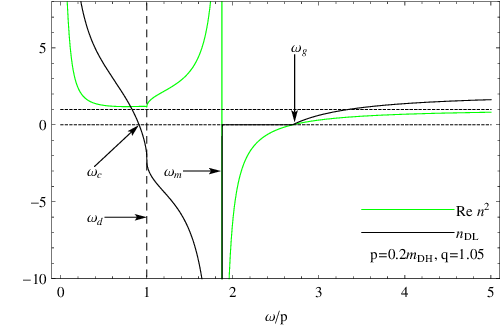}}
\end{minipage}}
\centerline{
\begin{minipage}[h]{0.48\textwidth}
\centering{\includegraphics[width=7.5cm,height=5cm] {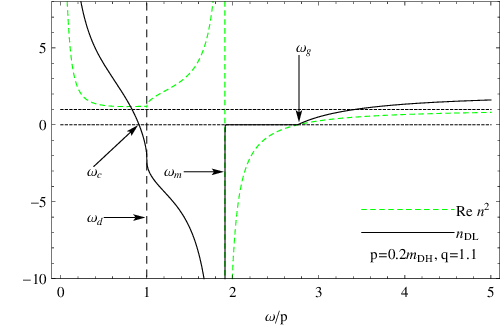}}
\end{minipage}
\begin{minipage}[h]{0.48\textwidth}
\centering{\includegraphics[width=7.5cm,height=5cm] {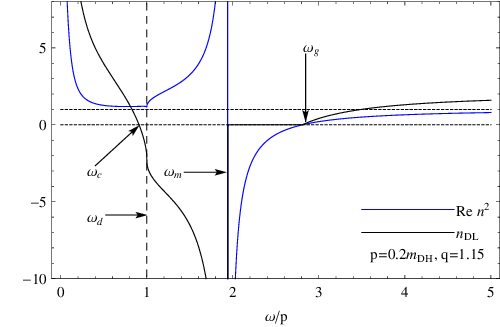}}
\end{minipage}}
\begin{minipage}[h]{0.48\textwidth}
\centering{\includegraphics[width=7.5cm,height=5cm] {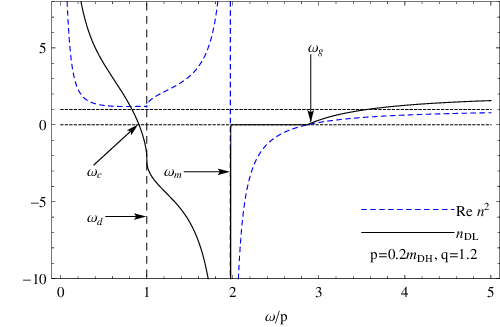}}
\end{minipage}
\caption{(color online). The Depine-Lakhtakia index $n_{DL}$ and the real part of $n^2$  of a nonextensive QGP.} \label{n2ndl}
\end{figure*}

We present ${\rm Re}\, n^2$ and the Depine-Lakhtakia index $n_{DL}$  in Fig.5  and in Fig.6 respectively.
Because $n^2$ and $n_{DL}$ are defined by  the electric permittivity and the  magnetic permeability,
 the properties of $n^2$ and $n_{DL}$ are determined by $\varepsilon$ and $\mu$.
${\rm Re}\, n^2$ and $n_{DL}$ demonstrate a singularity  at frequency $\omega_m$, which is just the pole of $\mu_M$.
When $q$ increases, as discussed in the last paragraph, the pole frequency  $\omega_m$  of ${\rm Re}\, n^2$ and $n_{DL}$ turns to large values.
${\rm Re}\, n^2$ and $n_{DL}$ have a frequency inflexion located at $\omega_d=p$ which is just the pole frequency of the real part of the electric permittivity. When $\omega<\omega_d$, $q$ has almost no any effect on ${\rm Re}\, n^2$ and $n_{DL}$, while for $\omega>\omega_d$, $q$ significantly influences ${\rm Re}\, n^2$ and $n_{DL}$.

We depict ${\rm Re}\, n^2$ and $n_{DL}$ in the same plots  in Fig.7 for different $q$ values for the case $p=0.2m_{DH}$. From Fig.6 and Fig.7, it is easy to see ${\rm Re}\, n^2$ and $n_{DL}$ intersect at frequency $\omega=\omega_g$.
In frequency regions $\omega<\omega_c$  and  $\omega>\omega_g$, $n_{DL}>0$, one can obtain a general refractive index. In the frequency region $\omega\in[\omega_m,\omega_g]$, $n_{DL}=0$ but ${\rm Re}\, n^2<0$ which implies a pure imaginary refractive index, electromagnetic wave does not propagate due to severe damping\cite{wang,jamal}. There is a  frequency region $\omega\in[\omega_c,\omega_m]$ in which $n_{DL}<0$. In terms of discussion in Sec.\ref{em},  the refractive index seems negative in that frequency region.
With the increase of $q$, the frequency range for $n_{DL}<0$ becomes wider. In addition, as shown from top to bottom panel in Fig.6, when the momentum $p$ increases, the frequency range for $n_{DL}<0$ becomes narrower.

The start point $\omega_c$ of the frequency region $\omega\in[\omega_c,\omega_m]$ for $n_{DL}<0$ is around the electric permittivity pole, while the  magnetic permeability pole $\omega_m$ determines the end point\cite{jiang13,jiang16}. As shown in Fig.6, when $\omega<\omega_d$, the several nonextensive curves of the Depine-Lakhtakia index $n_{DL}$ superpose each other, $q$ does not change the position of $\omega_c$.  While as $q$ increases, the pole frequency  $\omega_m$ of $\mu_M$ turns to large values. These facts lead to an enlargement of the frequency range for $n_{DL}<0$. When momentum $p$ increases, the change of momentum does not influence the position of $\omega_c$, as shown from top  to bottom panel in Fig.6. While the frequency $\omega_m$ of $\mu_M$  pole significantly reduces as momentum $p$ increases. As a result, the frequency range for $n_{DL}<0$ becomes narrower as the increase of the momentum.

\begin{figure}
\includegraphics[width=7.5cm,height=4.635cm] {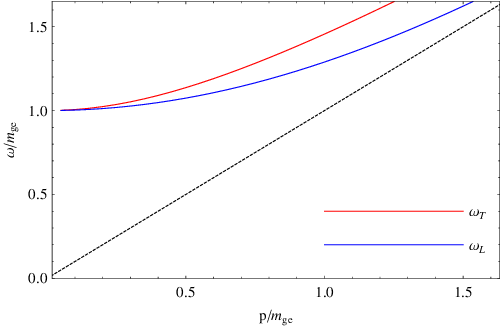}
\caption{(color online) The dispersion relations $\omega(p)$ for the longitudinal (blue curve) and transverse (red curve) gluons in unit of $m_{ge}=\frac{m_{DE}}{\sqrt{3}}$. The dotted line is the light cone $\omega=p$.} \label{effn}
\end{figure}

The criterion (\ref{neff}) $n_{DL}<0$ has been widely used to judge the existence of the negative refraction in a medium.
However, some researchers have argued that  besides that criterion, the propagating mode should satisfy the dispersion relation $n^2=\frac{p^2}{\omega^2}$  simultaneously\cite{wang}. In terms of the longitudinal and transverse gluon self-energies (\ref{logselfenergy})(\ref{transelfenergyfinal}), we can obtain the longitudinal and transverse dispersion relations by following formula ${\rm det}\,[p^2g^{\mu\nu}-p^\mu p^\nu-\Pi^{\mu\nu}]=0$\cite{thoma2000}. The solutions to dispersion relations are shown in Fig.8. As we can see in that figure, both the longitudinal and transverse gluons have one stable propagating mode above the light cone $\omega>p$ for $p>0$. Therefore, one should examine the potential propagating modes for the negative refractive index in the frequency region $\omega>p$.
It is instructive to
combine ${\rm Re} n^2$ and the Depine-Lakhtakia index $n_{DL}$ in the same plots in Fig.7 to perform the analysis.
In the frequency region $\omega\in[\omega_d, \omega_m]$, $\omega>p$ and $n_{DL}<0$.
While it  is clear that ${\rm Re}\, n^2>1$  in the same frequency region, as shown in plots in Fig.7. There are no solutions for $n^2=\frac{p^2}{\omega^2}$  in that frequency range, i.e., there are no propagating modes for the negative refractive index.

There exists not only temporal dispersion, but also spatial dispersion in a plasma.
Usually, the spatial dispersion is small in an isotropic medium. Therefore, researchers have studied the
frequency behavior of electromagnetic properties but omitted the spatial dispersion\cite{wang}.
However, as shown from Figure.1 to Figure.6, when momentum $p$ increases from $0.2m_{DH}$  to $0.3m_{DH}$, momentum $p$ significantly influences the electric permittivity, the magnetic permeability, ${\rm Re}\, n^2$ and $n_{DL}$. These facts indicate that the spatial dispersion is important in the electromagnetic responses of the QGP. As a matter of fact, though the authors in Ref.\cite{wang} have not investigated the effects of the spatial dispersion on the electromagnetic responses of the QGP explicitly,  the comparison of Fig1.($k=0$) with Fig.2. ($k=0.2m_D$) in that literature implies the importance of the spatial dispersion on the Depine-Lakhtakia index $n_{DL}$.

A relevant study to the present work also employed Eq.(\ref{poltensor0})  to investigate the chromo-refractive index of the hot quark-gluon plasma  medium, incorporating both medium effects and anisotropic effects\cite{jamal}. To account for the medium effects, the authors utilized the effective fugacity quasiparticle model (EQPM)\cite{eqmp}, where the effective fugacities $z_{g,q}$ describe the strong interaction effects within the medium.
As the nonextensive parameter $q$, the effective fugacities $z_{g,q}$ are embedded in the Debye mass.
The introduction of anisotropy   can be constructed from the isotropic distribution function $f_{iso}(\textbf{k}^2)$ by the rescaling of only one direction in momentum space  $f(\textbf{k})\sim f_{iso}(\textbf{k}^2+\xi(\textbf{k}\cdot\textbf{\^{n}})^2)$\cite{jamal,schenke,strickland2003,guoyun2023,aniso1,aniso2}.
However, the anisotropic distribution function $f(\textbf{k})$ does not satisfy Eq.(\ref{diffdis}), but
$\frac{\partial f(\textbf{k})}{\partial k^l}=\frac{v^l+\xi(\textbf{v}\cdot\textbf{\^{n}})n^l}{\sqrt{(1+\xi(\textbf{v}\cdot\textbf{\^{n}})^2)}}\frac{\partial f_{iso}(\textbf{k}^2)}{\partial k}$\cite{jamal,schenke,strickland2003,guoyun2023,aniso1,aniso2}. The anisotropic parameter $\xi$ will not be encoded  in Debye mass in the polarization tensor expression (\ref{poltensordeco}), but the integral over the solid angle $\Omega$, which will lead to complicated structures of  $\Pi^{ij}(P)$.
One has to introduce four projector operators to decompose anisotropic $\Pi^{ij}(P)$ into four structure functions\cite{jamal,schenke,strickland2003,guoyun2023,aniso1,aniso2}
\begin{eqnarray}
\Pi^{ij}=\alpha P^{ij}_T + \beta_L P^{ij}_L + \gamma P^{ij}_n + \delta P^{ij}_{kn}.
\end{eqnarray}

In an isotropic medium, $\Pi^{ij}(P)$ can be decomposed  into  two parts, \ie
the longitudinal and transverse gluon self-energies $\Pi_L$, $\Pi_T$. Electromagnetic wave interacts with medium in the same manner.
While in an anisotropic plasma medium, there are two polarization states and we can obtain two eigenvalues of electric permittivity in terms of the obtained four structure functions\cite{jamal}
\begin{eqnarray}
\varepsilon_L=1-\frac{\beta'_L}{P^2},\ \ \ \ \  \varepsilon_R=1-\frac{\beta'_L}{P^2}-\frac{\gamma}{P^2},
\end{eqnarray}
where $\beta'_L$ is related to $\beta_L$.
These results lead to  two different refractive indices denoted as
\begin{eqnarray}
n_{L/R}=\sqrt{\varepsilon_{L/R}\mu_M},
\end{eqnarray}
which is the so called birefringence or double refraction\cite{jamal}. During the early stage of the relativistic heavy ion collisions where the plasma will experience an anisotropic expansion in the momentum space,
or in off-central heavy ion collisions where the  strong magnetic field will be produced,  birefringence will be a common phenomenon in the medium, which might lead to some appreciable effects on some heavy ion observables.

\section{Summary}
\label{summary}
We have worked out the electric permittivity $\varepsilon(\omega,k)$ and the magnetic permeability $\mu_M(\omega,k)$ in terms of the nonextensive distribution functions  associated with the polarization tensor derived from kinetic theory, through which we have derived  the real part of the square of the refractive index ${\rm Re}\, n^2$ and the Depine-Lakhtakia index $n_{DL}$. We have investigated the effects of nonextensivity on the electromagnetic responses of the QGP. The real and imaginary parts of the electric permittivity have changed noticeably in the spacelike region as $q$ increases, but the pole frequency of the real part and inflexion frequency of the imaginary part remain unchanged. $\mu_M(\omega,k)$, ${\rm Re}\, n^2$ and $n_{DL}$ diverge at the frequency $\omega_m$ which shifts to large frequency region as the nonextensive parameter $q$ increases. There is a quite large frequency range $\omega\in[\omega_c,\omega_m]$ in which $n_{DL}<0$. And that frequency range becomes wider as $q$ increases.
Though $n_{DL}<0$ in the frequency region $\omega\in[\omega_c,\omega_m]$, there are no propagating modes  for the negative refractive index.

In addition, momentum $p$ reduces the real and imaginary parts of  the electric permittivity in the spacelike region. Whereas at the frequency pole of  the electric permittivity, the relation $\frac{\omega}{p}=1$ does not change as momentum $p$ varies. On the other hand, pole frequency of $\mu_M(\omega,k)$, ${\rm Re}\, n^2$ and $n_{DL}$ turns to small frequency region remarkably as momentum $p$ increases. These facts indicate that the spatial dispersion has  important effects on the electromagnetic responses of the QGP.

The refractive index can directly relate to  phenomenology of ultrarelativistic heavy-ion collisions.
Based on the refractive index at finite temperature or density, some groups have studied the medium responses due to the Cherenkov gluon radiation induced by the parton jet traveling through the hot QGP to explain the double peak structure of the experimental dihadron correlation function in the away side\cite{koch,dremin}.
By incorporating polarization and absorption effects with a complex index of refraction,
some authors have investigated the radiation energy loss of an energetic charge, which is helpful for understanding the jet quenching phenomena\cite{bluhm1}. In addition, some researchers have studied the effects of the QCD medium refraction on the elliptic flow and higher-order harmonics of prompt photons in ultrarelativistic heavy-ion collisions\cite{monnai}.

Generally, negative refraction of light will lead to some special medium properties, such as modified Snell's law, perfect lens, reversed Doppler shift, and so on\cite{ramakrishna}.  If the QGP has a negative refraction index,  the fast parton may induce unique Cherenkov radiation phenomenon in the medium.
Unlike the normal refraction medium, the radiation from the fast parton  is emitted in the opposite direction to the parton's motion direction\cite{ramakrishna}, which may cause unusual energy deposit and medium response.
Therefore, the  negative refraction  might result in some novel effect on jet quenching observables. In addition, the negative refraction might have an impact on the photon production
in ultrarelativistic heavy ion collisions.  These are  interesting issues deserving  further
investigations.

As discussed  in Ref.\cite{jiang16}  that besides the combination of the electric permittivity $\varepsilon$ and the magnetic permeability $\mu_M$, the longitudinal and transverse dielectric functions $\varepsilon_L,\varepsilon_T$ are usually used to describe the electromagnetic properties in plasma alternatively. The combination ($\varepsilon_L,\varepsilon_T$) can be related to ($\varepsilon$, $\mu_M$). Therefore, based on the gluon longitudinal and transverse self-energies incorporating the nonextensive effects,  $\varepsilon_L$  and  $\varepsilon_T$ can be derived, through which one can study nonextensive effects on diverse aspects of the QGP property, such as, Debye screening, Landau damping, dissociation of heavy quarkonia, photon and dilepton production, jet quenching and the related wakes. These are interesting issues which will be focused on in our future investigations.

{\bf Acknowledgment}
We are very grateful to Wen-chao Zhang, Hua Zheng and Li-lin Zhu for helpful discussion.
They drew us attention to Refs.\cite{wong,zheng0} which motivate the research.
This work is supported in part by the National Key Research and Development Program of China under Contract No. 2022YFA1604900. This work is also partly supported by the National Natural Science Foundation of China (NSFC) under Grant Nos.\ 12265013, 12435009 and 12275104 and Science and Technology Research Project of Education Department of Hubei Province under Grants No. D20221901.

\appendix
\section{A detailed derivation of Eq.(\ref{diffdis})}
For massless plasma constituents, $K=(E_k,\textbf{k})$ and $E_k=k=|\textbf{k}|=((k^1)^2+(k^2)^2+(k^3)^2)^\frac{1}{2}$,
the expanded nonextensive distribution function Eq.(\ref{appnondis}) can be expressed as
\begin{eqnarray}\label{a1}
n_{q,\bar{q},g}=\textbf{e}^{-\beta k}-(q-1)\beta k\textbf{e}^{-\beta k}+\frac{1}{2}(q-1)\beta^2 k^2\textbf{e}^{-\beta k}+...
\end{eqnarray}
By performing partial derivatives of $\textbf{e}^{-\beta k}$ over $k^l$ and $k$, one can obtain
\begin{eqnarray}\label{a2}
\frac{\partial (\textbf{e}^{-\beta k})}{\partial k^l}&=&\textbf{e}^{-\beta k}(-\beta)\frac{\partial k}{\partial k^l}\\ \nonumber&=&\textbf{e}^{-\beta k}(-\beta)\frac{\partial((k^1)^2+(k^2)^2+(k^3)^2)^\frac{1}{2} }{\partial k^l} \\ \nonumber
&=&(-\beta)\textbf{e}^{-\beta k}\cdot \frac{k^l}{k},
\end{eqnarray}
and
\begin{eqnarray}\label{a3}
\frac{\partial (\textbf{e}^{-\beta k})}{\partial k}&=&(-\beta)\textbf{e}^{-\beta k}.
\end{eqnarray}
In the same way, we can derive the corresponding partial derivatives of $k$ over $k^l$ and $k$
\begin{eqnarray}\label{a4}
\frac{\partial (k)}{\partial k^l}&=& \frac{k^l}{k},
\end{eqnarray}
and
\begin{eqnarray}\label{a5}
\frac{\partial (k)}{\partial k}&=&1.
\end{eqnarray}

According to Eqs.(\ref{a2})(\ref{a3})(\ref{a4}) and (\ref{a5}), from Eq.(\ref{a1}) one can get
\begin{equation}
\frac{\partial n_{q,\bar{q},g}}{\partial k^l}=\frac{k^l}{k}\frac{\partial n_{q,\bar{q},g}}{\partial k}=v^l\frac{\partial n_{q,\bar{q},g}}{\partial k}, \nonumber
\end{equation}
through which one can easily prove Eq.(\ref{diffdis}).
\section{A detailed derivation of the square of Debye mass Eq.(\ref{debyemasse})}
According to the expanded nonextensive distribution functions (\ref{appnondis}), one can obtain
\begin{eqnarray}\label{diffnonext}
\frac{\partial n_{q,\bar{q},g}}{\partial k}
&=&-\beta \textbf{e}^{-\beta k}+(q-1)\\ \nonumber &\cdot&(-\frac{1}{2}\beta^2k^2+2\beta k-1)\beta \textbf{e}^{-\beta k}.
\end{eqnarray}
In terms of the definition of the Debye mass (\ref{debyemass}),(\ref{dis})
and Eq.(\ref{diffnonext}), one can arrive the square of the Debye mass in nonextensive  hot QGP
\begin{widetext}
\begin{eqnarray}
m_{DE}^2&=&-\frac{g^2}{2\pi^2}\int_0^\infty k^2 dk \frac{\partial f(\textbf{k})}{\partial k}
\\ \nonumber&=&-\frac{g^2(N_c+N_f)}{\pi^2}\int_0^\infty k^2 \{-\beta \textbf{e}^{-\beta k}+(q-1)\cdot(-\frac{1}{2}\beta^2k^2+2\beta k-1)\beta \textbf{e}^{-\beta k}\}dk  \\ \nonumber
&=&\frac{g^2(N_c+N_f)}{\pi^2}\int_0^\infty k^2\beta \textbf{e}^{-\beta k}dk+
\frac{g^2(N_c+N_f)}{\pi^2}\int_0^\infty \frac{1}{2}(q-1)\beta^3k^4\textbf{e}^{-\beta k}dk
 \\ \nonumber
&-&  \frac{2g^2(N_c+N_f)}{\pi^2}\int_0^\infty (q-1)\beta^2k^3\textbf{e}^{-\beta k}dk+\frac{g^2(N_c+N_f)}{\pi^2}\int_0^\infty (q-1) k^2\beta \textbf{e}^{-\beta k}dk
\\ \nonumber
&=&\frac{2g^2T^2(N_c+N_f)}{\pi^2}+\frac{12g^2T^2(q-1)(N_c+N_f)}{\pi^2}-\frac{12g^2T^2(q-1)(N_c+N_f)}{\pi^2}+ \frac{2g^2T^2(N_c+N_f)(q-1)}{\pi^2}\\
&=&\frac{2(N_c+N_f)(1+(q-1))g^2T^2}{\pi^2}
\end{eqnarray}
\end{widetext}
which is just Eq.(\ref{debyemasse}).

\begin{widetext}
\section{A detailed derivation of the longitudinal and transverse self-energies}
According to Eq.(\ref{poltensordeco}) and longitudinal projector, we can get the structure function $\beta_L$
\begin{eqnarray}\label{gamma}
\beta_L&=&\frac{p^ip^j}{p^2}\Pi^{ij}(P)=m_{DE}^2 \frac{p^ip^j}{p^2} \int_0^\pi \frac{d\Omega}{4\pi}v^iv^l  [\delta^{jl}+\frac{p^lv^j}{\omega-\textbf{p}\cdot\textbf{\^v}+i0^+}] \\ \nonumber
&=&\frac{m_{DE}^2}{2}  \int_0^\pi sin\theta d\theta \frac{p^ip^j}{p^2}v^iv^l  [\delta^{jl}+\frac{p^lv^j}{\omega-\textbf{p}\cdot\textbf{\^v}+i0^+}]\\ \nonumber
&=&\frac{m_{DE}^2}{2} \int_0^\pi sin\theta d\theta [\frac{p^ip^jv^iv^l\delta^{jl}}{p^2}+ \frac{p^ip^jv^iv^lp^lv^j}{p^2}\cdot\frac{1}{\omega-\textbf{p}\cdot\textbf{\^v}+i0^+}]\\ \nonumber
&=&\frac{m_{DE}^2}{2}  \int_0^\pi sin\theta d\theta [cos^2\theta+\frac{pcos^3\theta}{\omega-pcos\theta+i0^+}]\\ \nonumber
&=&\frac{m_{DE}^2}{2}  \int_{-1}^1 dx [x^2+\frac{px^3}{\omega-px+i0^+}]\\ \nonumber
&=&-\frac{m_{DE}^2}{2}\cdot \frac{\omega^2(2p-\omega\cdot\ln[\frac{\omega+p+i0^+}{\omega-p+i0^+}])}{p^3}.
\end{eqnarray}
In terms of the structure function $\beta_L$ and Eq.(\ref{logselfenergy1}), we can obtain the longitudinal gluon self-energy
\begin{eqnarray}
\Pi_L(P)
=\frac{2(1+(q-1))(N_c+N_f)}{\pi^2}g^2T^2\cdot \frac{P^2}{p^2}\cdot [-1+\frac{\omega}{2p}(\ln|\frac{\omega+p}{\omega-p}|-i\pi\delta(p^2-\omega^2))].
\end{eqnarray}

Similarly, we can derive the transverse gluon self-energy by contracting  $\Pi^{ij}$ with respect to the transverse projector
\begin{eqnarray}
\Pi_T(P)&=&\frac{1}{2}(\delta^{ij}-\frac{p^ip^j}{p^2})\Pi^{ij}(P)=\frac{m_{DE}^2}{2}  \int_0^\pi \frac{d\Omega}{4\pi} (\delta^{ij}-\frac{p^ip^j}{p^2}) v^iv^l  [\delta^{jl}+\frac{p^lv^j}{\omega-\textbf{p}\cdot\textbf{\^v}+i0^+}]\\ \nonumber
&=&\frac{m_{DE}^2}{4}  \int_0^\pi sin\theta d\theta [v^jv^j-\frac{p^ip^jv^iv^j }{p^2}+\frac{p^lv^jv^jv^l}{\omega-\textbf{p}\cdot\textbf{\^v}+i0^+}-\frac{p^lv^jp^ip^jv^iv^l}{p^2(\omega-\textbf{p}
\cdot\textbf{\^v}+i0^+)}]\\ \nonumber
&=&\frac{m_{DE}^2}{4}  \int_0^\pi sin\theta d\theta [1-cos^2\theta +\frac{pcos\theta}{\omega-pcos\theta+i0^+}-\frac{pcos^3\theta}{\omega-pcos\theta+i0^+}]\\ \nonumber
&=&\frac{m_{DE}^2}{4}  \int_{-1}^1  dx [1-x^2 +\frac{px}{\omega-px+i0^+}-\frac{px^3}{\omega-px+i0^+}]\\ \nonumber
&=&\frac{m_{DE}^2}{4}  \int_{-1}^1  dx [1 +\frac{px}{\omega-px+i0^+}]-\frac{m_{DE}^2}{4}  \int_{-1}^1  dx [x^2 +\frac{px^3}{\omega-px+i0^+}]\\ \nonumber
&=&\frac{m_{DE}^2}{4}  \int_{-1}^1  dx [1 +\frac{px}{\omega-px+i0^+}]-\frac{\beta_L}{2}\\ \nonumber
&=&\frac{m_{DE}^2}{4} \frac{\omega}{p}\ln[\frac{\omega+p+i0^+}{\omega-p+i0^+}]+\frac{m_{DE}^2}{4}\cdot \frac{\omega^2(2p-\omega\cdot\ln[\frac{\omega+p+i0^+}{\omega-p+i0^+}])}{p^3}\\ \nonumber
&=&\frac{(1+(q-1))(N_c+N_f)}{\pi^2}g^2T^2 \cdot[\frac{\omega^2}{p^2}+\frac{\omega(p^2-\omega^2)}{2p^3}\cdot(\ln|\frac{\omega+p}{\omega-p}|-i\pi\delta(p^2-\omega^2))]
\end{eqnarray}
\end{widetext}

\end{document}